\documentclass[twocolumn,prb,aps,showpacs]{revtex4}

\usepackage{graphicx}
\def\be{\begin{equation}}
\def\ee{\end{equation}}
\def\bea{\begin{eqnarray}}
\def\eea{\end{eqnarray}}
\def\ba{\begin{array}}
\def\ea{\end{array}}
\def\bdm{\begin{displaymath}}
\def\edm{\end{displaymath}}

\begin{document}

\title{Theory of Fermionic superfluid with SU(2)$\times$SU(6) symmetry}

\author{S.-K. Yip}

\affiliation{Institute of Physics, Academia Sinica, Nankang, Taipei
115, Taiwan}

\date{\today }

\begin{abstract}

We study theoretically interspecies Cooper pairing in a fermionic system with SU(2)$\times$SU(6)
symmetry.  We show that, with suitable
unitary transformations,  the order parameter for the ground
state can be reduced to only two non-vanishing complex components.
The ground state has a large degeneracy.  We find that while some
Goldstone modes have linear dispersions, others are quadratic at low frequencies.
We compare our results with the case of SU(N).

\end{abstract}



\maketitle

\section{Introduction}\label{Intro}

Higher symmetry groups such as SU(3) play an essential role in our understanding
of elementary particle physics \cite{Georgi}.  However, in typical condensed matter
systems, the internal symmetries are much simpler.  Superfluid $^4$He has no spin,
whereas $^3$He and electrons in solids have only spin $1/2$.  The situation changes
with the advances in cold-atomic gases.  We have already seen many examples of
bosonic systems with finite spins ($\ge 1$) \cite{sb}.  There has also been much
attention in fermionic systems with more complex internal structure
\cite{sf}, and more recently, in systems where the symmetry is higher than the usual
spin rotational symmetries.  Examples include the hidden symmetry
in spin $3/2$ fermions \cite{3/2}, approximate SU(3) symmetry for $^6$Li
near special external magnetic field values \cite{su3-ex},
\cite{Modawi97,HH,He06,Cherng07,CY08,Martikainen09,Ozawa10,su3-t},
and the enlarged symmetry for atoms with finite nuclear but no net electronic spin
\cite{Yb,Sr,su-n-t}.

With no net electronic spin, the hyperfine spin of an atom comes
entirely from its nucleus.
Since the interaction between atoms mainly arises from their electronic clouds, the
interatomic interaction is then independent of the total spins of the
atoms involved.
(This is in contrast with the more general situation where
the scattering length between two atoms in general depends on their total
spin.  See ref \cite{sb,sf}).
A very interesting system of this class has been studied recently experimentally,
namely a mixture of $^{171}$Yb and $^{173}$Yb \cite{Taie}.
$^{171}$Yb and $^{173}$Yb have nuclear spins $1/2$ and $5/2$ respectively.
 For the $^{171}$Yb and $^{173}$Yb mixture under
discussion, the intra-species interaction is weak
(the s-wave scattering length  is $-0.15 nm$ between two $^{171}$Yb atoms
and $10.55nm$ between two $^{173}$Yb atoms).  However, there is
a rather large inter-species attraction, with the scattering length
$a \equiv a_{171-173} \approx - 30.6 nm$ \cite{Taie}.  The Kyoto group has already been able
to cool this mixture much below the degeneracy temperature, thus raising
the interesting possibility of interspecies Cooper pairing in this system.

Motivated by this, we study a two species fermionic system with interspecies
attractive interaction.  The weak intraspecies interaction is expected to only
slightly modify the quantitative details of the system and will be ignored.
We shall take a SU(2) internal symmetry for the first species, and SU(6)
for the second one, though our findings are immediately generalizable to
SU($2f+1$) with general half-integer $f$'s.  The system is expected to
undergo interspecies Cooper pairing.  (We shall  only consider weak
attractive interactions, and therefore ignore possibilities of
bound states involving three or more particles, {\it c.f.}
\cite{su3-ex,su3-t}).  The general  order parameter is
thus a $2 \times (2f+1)$ complex valued matrix, since pairing can occur
between any internal state of the first species and that of the second species.
We determine the structure of the order parameter for the ground state,
employing the mean-field approximation.  We show that,
with a suitable choice of basis,  the order parameter for the ground
state can be reduced to only two non-vanishing complex components.
The ground state for this system has also investigated before \cite{Dickerscheid}
in the case where the interaction depends on the total spin $F$ of the
two interacting atoms and therefore the Cooper pairs have a definitive spin $F$.
We shall mention briefly the relation of our results to theirs.

The ground state is found to possess a large degeneracy.  We thus proceed
to find the Goldstone modes of this system.  We show that there are
$2 \times (2f+1)$ such modes, and we shall evaluate their dispersion relations.
We find $4$ linear modes, and $2 \times (2 f -1 )$ modes which have
quadratic dispersion at very low frequencies, but becoming also
linear as slightly higher frequencies.  We also determine
 the physical variables coupling to each of these modes,
thus indicating how these modes can be excited experimentally.

   Currently,
   the fermi temperature $T_F$ in the experiment of \cite{Taie} is around $200$nK, thus the
   product $|k_F a| \approx 0.36$ (here $k_F$ is the fermi wavevector).
     While this allows
   a weak-coupling consideration as in here, the transition temperature
   is unfortunately low ($T_c/T_F \sim e^{ - \frac{\pi}{|2 k_F a|}} \sim 0.01$).
     However, we hope that eventually the
   superfluid state can be reached (perhaps using
   optical Feshbach resonances to enhance the interaction \cite{Taie})
   and the physics discussed here be studied.
   Also, we shall see that many of our physical results are more general, and thus
   would be applicable in case other more favorable related systems can be found.

This paper is organized as follows.  In section \ref{G} we consider the
ground state properties, and in section \ref{Modes} we discuss the collective
modes.  Our  results have many similarities but some differences
with the more studied case of SU(N) fermi superfluids.  We therefore
compare our results with this case in Section \ref{Comp}.
We conclude in section \ref{Concl}.

\section{Ground state}\label{G}

In this section we discuss the ground state properties,
assuming mean-field theory.  This requires that
the scattering length between the $^{171}$Yb and $^{173}$Yb atoms
be small compared with the interparticle distances, and also
that the transition temperature is small compared  with
the fermi temperatures.  Nevertheless, in below we shall argue that
many of our results are qualitatively correct beyond mean-field
approximations.

Let us denote the annihilation operators for the
$^{171}$Yb and $^{173}$Yb atoms by $a_{\vec k, \lambda}$ and
$c_{\vec k, \nu}$ respectively, where $\vec k$ is the
wavevector and $\lambda = \pm \frac{1}{2}$, $\nu = - f, ..., f$
denotes the internal states.  The Hamiltonian $H$ has two parts.
The kinetic energy is given by
\be
H_{K} = \sum_{\vec k, \lambda} \xi_k
   a^{\dagger}_{\vec k, \lambda} a_{\vec k, \lambda}
    + \sum_{\vec k, \nu} \xi_k
   c^{\dagger}_{\vec k, \nu} c_{\vec k, \nu}
   \label{K}
\ee
Here $\xi_k \equiv \frac{k^2}{2m} - \mu$,
 $m$ is the mass of the atoms
 (We ignore the small mass difference between $^{171}$Yb and $^{173}$Yb here)
For simplicity we shall also confine ourselves to the case
where the chemical potentials $\mu$ of the two species to be equal
(and there are no chemical potential differences among the different
hyperfine sublevels)\cite{pop}.
The interaction $H_{int}$ is given by

\be
H_{int} = g \sum_{\vec k, \vec k', \vec q, \lambda, \nu}
a^{\dagger}_{\vec k_+, \lambda} c^{\dagger}_{-\vec k_-, \nu}
c_{-\vec k'_-, \nu} a_{\vec k'_+, \lambda}
\label{Hint}
\ee
where $\vec k_{\pm} \equiv \vec k \pm \vec q/2$.
This is the most general interspecies s-wave interaction obeying
SU(2)$\times$SU(6) symmetry.
$g$ can be eliminated in favor of the scattering length
$a_{171-173}$ but we shall not need this
explicit relation here.

Within mean-field theory we can replace $H_{int}$ by an effective interaction
\be
H_{int}^{eff} = \sum_{\lambda,\nu}
\left\{ \sum_{\vec k} \left( \Delta_{\lambda,\nu}
a^{\dagger}_{\vec k, \lambda} c^{\dagger}_{-\vec k,\nu} +
\Delta^*_{\lambda,\nu} c_{-\vec k, \nu} a_{\vec k, \lambda} \right)
  - \frac{|\Delta_{\lambda,\nu}|^2}{g} \right\}
\label{mf}
\ee
where $\Delta_{\lambda,\nu}$ has to satisfy the self-consistent equation
\be
\Delta_{\lambda,\nu} = (-g) \sum_{\vec k} < a_{\vec k, \lambda} c_{-\vec k, \nu} >
\label{self}
\ee

Before solving this mean-field Hamiltonian we apply the Ginzburg-Landau (GL) theory.
The GL free energy has the form
\be
\Omega = \alpha {\rm Tr} \left[ {\bf \Delta \Delta^{\dagger}} \right]
  + \sum_{ l \ge 2} \frac{\beta_l}{l} {\rm Tr} \left[ ({\bf \Delta \Delta^{\dagger}})^l \right]
\label{GL}
\ee
Here ${\bf \Delta}$ is a $2 \times (2f+1)$ matrix with elements $\Delta_{\lambda,\nu}$.
$\alpha$, $\beta_{l}$ can easily be evaluated but for here it is sufficient to
know that all $\beta_{l} > 0$, and $\alpha$ is positive above some transition temperature
$T_c$ and negative below it.  We would like to find the form of ${\bf \Delta}$ which
minimizes $\Omega$ below $T_c$.  For this, we notice that since all $\beta_l > 0$,
${\bf \Delta}$ must be such that these higher order terms are minimized for any given
${\rm Tr} \left[ {\bf \Delta \Delta^{\dagger}} \right]$.
Let us denote $D_{\lambda} \equiv [ \sum_{\nu} |\Delta_{\lambda,\nu}|^2 ]^{1/2}$.
We obtain
${\bf \Delta \Delta^{\dagger}} = \frac{D^2_{1/2}+ D^2_{-1/2}}{2}
  + {\bf M}$
where ${\bf M}$ is a Hermitian matrix, ${\rm Tr} [{\bf M}] = 0$, and when
expanded as ${\bf M} = M_1 \sigma_1 + M_2 \sigma_2 + M_3 \sigma_3$
 using Pauli matrices $\sigma_{1,2,3}$, we have
 $M_3 =  (D^2_{{1}/{2}} - D^2_{-{1}/{2}})/2$,
 $M_1 - i M_2 = \sum_{\nu} \Delta_{\frac{1}{2},\nu} \Delta^*_{-\frac{1}{2},\nu}$.
 We thus have ${\rm Tr} \left[ {\bf \Delta \Delta^{\dagger}} \right] =
 D^2_{{1}/{2}}+ D^2_{-{1}/{2}}$.  We also find easily
 ${\rm Tr} \left[ ({\bf \Delta \Delta^{\dagger}})^2 \right] =
  (D^2_{{1}/{2}}+ D^2_{-{1}/{2}})^2/2 + {\rm Tr} [{\bf M}^2]$.
 Thus, for given ${\rm Tr} \left[ {\bf \Delta \Delta^{\dagger}} \right]$,
 this fourth order term would be minimized
 if  we choose
 \be
 \sum_{\nu} \Delta_{\frac{1}{2},\nu} \Delta^*_{-\frac{1}{2},\nu} = 0
 \label{ortho}
 \ee
 and
 \be
 D_{{1}/{2}} = D_{-{1}/{2}} \equiv D  \ .
 \label{eD}
 \ee
 Hence, if we regard $\Delta_{{1}/{2},\nu}$ and
 $\Delta_{-{1}/{2},\nu}$ each as a (un-normalized) wavefunction of a spin $f$ particle,
 then eq (\ref{ortho}) requires that  these two wavefunctions are orthogonal,
 whereas eq (\ref{eD}) shows that they are of equal  magnitude.
 A possible choice of ${\bf \Delta}$ satisfying
 eq (\ref{ortho}) is one where all elements $\Delta_{\lambda,\nu}$ vanish
 except $\Delta_{1/2,1/2}$ and $\Delta_{-1/2,-1/2}$.  Eq (\ref{eD}) then
 requires that their magnitude to be equal.  We shall
 give an alternate explanation of eq (\ref{ortho}) and (\ref{eD}) below.\cite{D2}
 Using  similar  reasoning as above, we can actually see that in  fact all $l \ge 2$ terms
 are minimized by the conditions eq (\ref{ortho}) and (\ref{eD}).
 The free energy then becomes
 $\Omega = 2 (\alpha D^2 + \beta_2 D^4/2 + ...)$, the same as the usual BCS theory
 for a two-component system except an overall extra factor of $2$.

Now we return to the microscopic theory.  Defining $\Psi_{\lambda,\nu}$ via
$\Delta_{\lambda,\nu} = D_{\lambda} \Psi_{\lambda,\nu}$ thus
$\sum_{\nu} |\Psi_{\lambda,\nu}|^2 = 1$, the pairing term in $H_{int}^{eff}$ can
be written as $\sum_{\lambda} D_{\lambda} a^{\dagger}_{\vec k, \lambda}
\left[ \sum_{\nu} \Psi_{\lambda,\nu} c^{\dagger}_{-\vec k,\nu} \right] + {\it h.c.}$, and thus can be interpreted as pairing between
$a^{\dagger}_{\vec k, \lambda}$ and the state ${\tilde c}^{\dagger}_{-\vec k, \lambda}$
$\equiv \sum_{\nu} \Psi_{\lambda,\nu} c^{\dagger}_{-\vec k,\nu}$,
$\lambda = \pm \frac{1}{2}$.
(Here, for convenience of writing, we are simply calling the particles
 by their corresponding operators).
 Eq (\ref{ortho}) implies that the most favorable
state is such that ${\tilde c}^{\dagger}_{-\vec k,1/2}$ and
${\tilde c}^{\dagger}_{-\vec k,-1/2}$ are orthogonal to each other.
This is physical reasonable, as then $a^{\dagger}_{\vec k,1/2}$ and
$a^{\dagger}_{\vec k,-1/2}$ does not have to compete with each other to
pair with the c-atoms.  With ${\tilde c}^{\dagger}_{-\vec k,\pm 1/2}$
already defined above, we can introduce
${\tilde c}^{\dagger}_{-\vec k,\nu}$ for $\nu \ne \pm \frac{1}{2}$
to make a complete set, therefore a unitary transformation between
${\tilde c}^{\dagger}_{-\vec k,\nu}$ and ${c}^{\dagger}_{-\vec k,\nu}$ operators:
\be
{\tilde c}^{\dagger}_{-\vec k,\nu} = U_{\nu,\nu'} {c}^{\dagger}_{-\vec k,\nu'}
\label{transf}
\ee
where $U_{\nu,\nu'} = \Psi_{\lambda=\nu,\nu'}$ for $\nu = \pm \frac{1}{2}$,
and $U_{\nu,\nu'}$ for $\nu \ne \pm \frac{1}{2}$ can be arbitrary so long
as the matrix ${\bf U}$ is unitary.  The kinetic energy can be re-written
as $K = \sum_{\vec k, \lambda} \xi_k
   a^{\dagger}_{\vec k, \lambda} a_{\vec k, \lambda}
    + \sum_{\vec k, \nu} \xi_k
   {\tilde c}^{\dagger}_{\vec k, \nu} {\tilde c}_{\vec k, \nu}$,
   and $H_{int}^{eff}$ can now be written as
   $ \sum_{\lambda} \left[ D_{\lambda} ( a^{\dagger}_{\vec k,\lambda}
   {\tilde c}^{\dagger}_{-\vec k,\lambda} +
    {\tilde c}_{-\vec k, \lambda} a_{\vec k,\lambda} ) -
    \frac{|D_{\lambda}|^2}{g} \right]$.  Thus the pairing
    term is of the normal BCS form except the sum over $\lambda$.  Thus the
    standard BCS calculations can be immediately applied.  The quasiparticle
    energies are thus $E_{\vec k,\lambda} = \left[\xi_k^2 + D_{\lambda}^2 \right]^{1/2}$
    with $D_{\lambda}$ playing the role of the energy gap for the $\lambda$ species.
    For the expectation values, we have (for zero temperature, to which we confine ourselves
    for the rest of the paper)
    $< a_{\vec k, \lambda} {\tilde c}_{-\vec k, \lambda} > =
    \frac{D_{\lambda}}{2 E_{\vec k,\lambda}}$,
    $< a^{\dagger}_{\vec k,\lambda} a_{\vec k,\lambda} >
      = < {\tilde c}^{\dagger}_{\vec k,\lambda} {\tilde c}_{\vec k, \lambda}>$
      $\equiv n^0(E_{k,\lambda}) = \frac{1}{2}
      \left( 1 - \frac{\xi_k}{E_{k,\lambda}} \right)$ etc for $\lambda = \pm \frac{1}{2}$.
      We also have $< a_{\vec k, \lambda} {\tilde c}_{-\vec k, \nu} > = 0$
      whenever $\nu  \ne \lambda$.  Expectation values involving the $c$ operators
      can be obtained by the inverse transformation.  We get
      $< a_{\vec k, \lambda} { c}_{-\vec k, \nu} > $ $= $ $\sum_{\nu'} U_{\nu',\nu}
      < a_{\vec k, \lambda} {\tilde c}_{-\vec k, \nu'} > $ $=$
      $U_{\lambda,\nu} < a_{\vec k, \lambda} {\tilde c}_{-\vec k, \lambda} >$ $= $
      $\Psi_{\lambda, \nu} \frac{D_{\lambda}}{2 E_{k,\lambda}}$ $=$
      $\frac{\Delta_{\lambda,\nu}}{2 E_{k,\lambda}}$.
      The self-consistent equation eq (\ref{self}) becomes
      $\Delta_{\lambda,\nu} = (-g) \sum_{\vec k} \frac{\Delta_{\lambda,\nu}}{2 E_{k,\lambda}}$
      and thus, after multiplying by $\Delta^*_{\lambda,\nu}$ and sum over $\nu$,
      either $D_{\lambda} = 0$, or
      \be
      1 = (-g) \sum_{\vec k} \frac{1}{2 (\xi_k^2 + D_{\lambda}^2)^{1/2}}
      \label{gap-eq}
      \ee
      Thus $D_{\lambda}$ obeys the usual BCS gap equation.  From the form of
      the Hamiltonian, it is obvious that the most favorable state would have
      both $D_{1/2}$ and $D_{-1/2}$ finite, and by eq (\ref{gap-eq}), both
      attain the usual BCS value and thus equal, consistent with eq (\ref{eD}).
      The particles ${\tilde c}^{\dagger}_{\vec k,\nu}$ for $\nu \ne \pm \frac{1}{2}$
      are not involved in pairing.  They maintain the normal state energies
      $\xi_k$, and there are thus $(2f -1)$ remaining fermi surfaces,
      and $< {\tilde c}^{\dagger}_{\vec k,\nu} {\tilde c}_{\vec k,\nu} >
       = f(\xi_k)$, the fermi function.

       The above can be readily generalized to higher symmetries.  For example,
       for SU($4$) $\times$ SU($2f+1$) with $f \ge 3/2$, then the ground state
       has an order parameter which, in a suitable basis, can be reduced to pairing only
       between $a_{\lambda}$ and $c_{\lambda}$ with the same $\lambda$.
       There are $(2 f + 1) - 4 = (2 f  -3 )$ fermi surfaces remaining normal.

       \section{Collective Modes}\label{Modes}

       \subsection{Dispersion relations: weak pairing limit}\label{weakpairing}

       The ground state therefore has a very high degeneracy.  Any choice of
       the unitary transformation ${\bf U}$ gives identical ground state energy.
       The system is thus characterized by a large number of Goldstone modes,
       with the mode frequency $\omega$ vanishing as the wavevector $\vec q$
       approaches zero.  These modes are associated with the fluctuations of
       the order parameter components $\delta \Delta_{\lambda,\nu} (\vec q)$ away from their
       equilibrium values.  We shall employ the kinetic equation approach \cite{ke}
       to evaluate their dispersion.  This method is equivalent to the random phase
       approximation in diagrammatic approaches.  Though we would employ the
       weak-coupling approximation, we shall argue that many of our results
       remain qualitatively valid for general interaction strengths,
       provided that the broken symmetries for the ground state remain
       the same as that found within the weak-coupling approximation.
       For simplicity we shall restrict ourselves to zero temperature.

       To simplify our notation we shall
       drop the tildes on the ${\tilde c}$ operators, or equivalently take a reference
       state where ${\bf U}$ is the identity matrix.  Physical variables can be
       obtained easily by applying the unitary transformation ${\bf U}$.
       We list the different types of modes in turn:

       ({\it case 1}): Modes corresponding to $\delta \Delta_{\lambda,\nu}$ with $\nu \ne \pm \frac{1}{2}$:
        For definiteness we consider $\delta \Delta_{1/2,3/2} (\vec q)$.  Besides the
        mean-field pairing terms in eq (\ref{mf}), we include
        \be
        \delta H = \sum_{\vec k} \delta \Delta_{1/2,3/2}(\vec q)
        a^{\dagger}_{\vec k_+,1/2} c^{\dagger}_{-\vec k_-,3/2} + {\it h.c.}
        \label{dH}
        \ee
        where $\vec k_{\pm} \equiv \vec k \pm \frac{\vec q}{2}$.  The hermitian conjugate
        ({\it h.c.})
        term, involving $\delta \Delta^*_{1/2,3/2}$ and other $\delta \Delta_{\lambda,\nu}$
        turns out to be decoupled from the equations below.
        $\delta \Delta_{1/2,3/2} (\vec q)$ has to obey the self-consistent equation
        \be
        \delta \Delta_{1/2,3/2} (\vec q) = (-g) \sum_{\vec k}
        < a_{\vec k_+,1/2} c_{-\vec k_-,3/2} >^{(1)}
        \label{sc1}
        \ee
        where the superscript $^{(1)}$ denotes the first order fluctuation contribution.
        Its equation of motion can be easily obtained using the hamiltonian
        $H = H_{K} + H_{int}^{eff} + \delta H$.  We get
        \begin{widetext}
        \bea
        i \frac{\partial}{\partial t} < a_{\vec k_+,1/2} c_{-\vec k_-,3/2} >^{(1)}
          &=& \left(\xi_{k_+} + \xi_{k_-} \right)
          < a_{\vec k_+,1/2} c_{-\vec k_-,3/2} >^{(1)} +  \Delta_{1/2,1/2}
          < c^{\dagger}_{-\vec k_+,1/2} c_{-\vec k_-,3/2} >^{(1)} \nonumber \\
          & & \qquad + \delta \Delta_{1/2,3/2} (\vec q)
           \left( < c^{\dagger}_{-\vec k_-,3/2} c_{-\vec k_-,3/2} >^{(0)}  -
             < a_{\vec k_+, 1/2} a^{\dagger}_{\vec k_+, 1/2}>^{(0)} \right)
             \label{em1}
        \eea
        Here
        the superscript $^{(0)}$ stands for equilibrium expectation values.  Thus
        we need also the equation of motion for
        $  < c^{\dagger}_{-\vec k_+,1/2} c_{-\vec k_-,3/2} >^{(1)} $:
        \bea
        i \frac{\partial}{\partial t}
         < c^{\dagger}_{-\vec k_+,1/2} c_{-\vec k_-,3/2} >^{(1)}
            &=& - \left(\xi_{k_+} - \xi_{k_-} \right) < c^{\dagger}_{-\vec k_+,1/2} c_{-\vec k_-,3/2} >^{(1)}
            + \Delta^*_{1/2,1/2} < a_{\vec k_+,1/2} c_{-\vec k_-,3/2} >^{(1)} \nonumber \\
             & & \qquad
              - \delta \Delta_{1/2,3/2} (\vec q)
              < c^{\dagger}_{-\vec k_+,1/2} a^{\dagger}_{\vec k_+,1/2} >^{(0)}
       \label{em2}
       \eea
       obtaining thus a closed set of equations for
       $< a_{\vec k_+,1/2} c_{-\vec k_-,3/2} >^{(1)}$ and
       $< c^{\dagger}_{-\vec k_+,1/2} c_{-\vec k_-,3/2} >^{(1)} $.
       Fourier transform and solving for $< a_{\vec k_+,1/2} c_{-\vec k_-,3/2} >^{(1)}$
       and inserting into eq (\ref{sc1}),
       we get
       \be
       0 = \sum_{\vec k} \left\{ \left[
       \frac{ \frac{1}{2} \left( 1 - \frac{\xi_{k_+}}{E_{k_+}} \right) f(\xi_{k_-})}
        { \omega - \xi_{k_-} + E_{k_+} }
        - \frac{ \frac{1}{2} \left( 1 + \frac{\xi_{k_+}}{E_{k_+}} \right) ( 1 - f(\xi_{k_-}))}
        { \omega - \xi_{k_-} - E_{k_+} } \right]
        - \frac{1}{2 E_k} \right\}
        \label{d1}
        \ee
      where we have also used eq (\ref{gap-eq}) to eliminate the coupling constant $g$.
      The term in the square bracket is a pair susceptibility.
      We can also directly obtain it from evaluating the response function of
       $< a_{\vec k_+,1/2} c_{-\vec k_-,3/2} >^{(1)}$ to $\delta H$. The second term is
      for adding to the ground state a pair of $c$ and $a$ particles with energies
      $\xi_{k-}$ and $E_{k_+}$.  This process occurs
      only when the final states are available, hence the factor
      $ \frac{1}{2} \left( 1 + \frac{\xi_{k_+}}{E_{k_+}} \right) ( 1 - f(\xi_{k_-}))$
      in the numerator.  Similar interpretation applies to the first term which
      stands for annihilation of the pair.  Note that the fermi factor restricts
      $\xi_{k_-} < 0$ for this term, hence $E_{k_+} - \xi_{k_-} = E_{k_+} + |\xi_{k_-}| > 0$.
      We also note that eq (\ref{d1}) converges in the ultraviolet. \cite{convg}

      One can easily check that, if $\vec q=0$, eq (\ref{d1}) is satisfied with $\omega = 0$,
      showing that we indeed has a Goldstone mode.  We next search for a solution
      to eq (\ref{d1}) for small $q$ and $\omega$.  For this, we add to eq (\ref{d1})
      the vanishing quantity $ 0 = \sum_{\vec k}
      \left( - \frac{ f(\xi_{k_-})}{ 2 E_{k_+}} - \frac{ 1- f(\xi_{k_-})}{ 2 E_{k_+}}
       + \frac{1}{ 2 E_{k_+}} \right)$, and replace the dummy variable $\vec k$
       by $\vec k_{+}$, so that $\vec k_- \to \vec k$.  Expanding the resulting equations
       in $q$ and $\omega$, taking angular average, we obtain
       \be
       0 = A_1 \omega + A_2 \omega^2 + ( \frac{A_1}{2} + B_1 + B_2 ) \frac{q^2}{m} \label{d2}
       \ee
        where
       $A_1 \equiv \sum_{\vec k} \frac{1}{ 2 E_k}
       \left[  \frac{ 1- f(\xi_{k})}{ ( E_k +\xi_k)}
          - \frac{ f(\xi_{k})}{( E_k - \xi_k) }  \right]$,
      $A_2 \equiv \sum_{\vec k} \frac{1}{ 2 E_k}
       \left[  \frac{ 1- f(\xi_{k})}{ ( E_k +\xi_k)^2}
          + \frac{ f(\xi_{k})}{( E_k - \xi_k)^2 }  \right]$,
       $ B_1 \equiv - \frac{2}{3} \sum_{\vec k} \left( \frac{k^2}{2m} \right)
         \frac{ \xi_k }{ 2 E_k^2}
         \left[ \frac{ 1- f(\xi_{k})}{(E_{k}+ \xi_k) }
         \left( \frac{1}{E_k +\xi_k} + \frac{1}{E_k} \right)
          - \frac{ f(\xi_{k})}{ ( E_k - \xi_k) } \left( \frac{1}{E_k - \xi_k} + \frac{1}{E_k} \right)
          \right] $, and
          $ B_2 \equiv \frac{3}{8} \sum_{\vec k} \left[
          - \frac{ \frac{\xi_k}{2} + \frac{k^2}{6m}}{E_k^3} +
          \frac{ {\xi_k}^2 \frac{k^2}{2m}}{E_k^5} \right]$

    Formally the term $A_2 \omega^2$ is small and thus can be dropped in the $\omega \to 0$ limit.
    However, we shall see that in the weak-coupling regime, the coefficient $A_1$ is small, and hence
    this term need to be kept in general beyond some small frequency regime $\omega^*$ which
    we shall define later.

      Before we discuss this weak-pairing limit in which we are principally interested, it is instructive
      to evaluate first the dispersion in the strong pairing limit, where $\mu < 0$ and $|\mu| \gg \Delta$
      ({\it c.f.} \cite{Engelbrecht}).
      Here $\Delta$ stands for the value of $|\Delta_{1/2,1/2}| = |\Delta_{-1/2,-1/2}|$ in equilibrium.
      In this limit, we get $B_1 = -\frac{3}{4} A_1$, while $B_2$ is negligible. $A_2 = A_1 /(8 |\mu|)$
      and hence its contribution is always negligible when $\omega \ll |\mu|$. Hence
      we obtain $\omega = \frac{q^2}{4m}$, the energy of a free particle of mass $2m$
      (see also discussions below).

      Now we return to the weak-pairing limit, where $\mu \gg \Delta$.  The expression $A_1$ is explicitly
      particle-hole asymmetric, {\it i.e.}, if we approximate the density of states near the
      Fermi surface by a constant $N(0)$ and replace $\sum_{\vec k} \to N(0) \int d \xi$, $A$
      would vanish.  Hence we must use the more accurate expression
      $\sum_{\vec k} \to \frac{m (2 m \mu)^{1/2}}{2 \pi^2} \int_{-\mu}^{\infty} d\xi
      \left( 1 + \frac{\xi}{\mu} \right)^{1/2} $.  Dividing the region of integration to
      $|\xi|< \mu$ and $|\xi| > \mu$, one can show that the later is smaller than the former
      in the $\mu \gg \Delta$ limit.  The first contribution to $A_1$ can be re-written
      as $ \frac{m (2 m \mu)^{1/2}}{2 \pi^2} \int_{0}^{\mu} d\xi \left[
      \left( 1 + \frac{\xi}{\mu} \right)^{1/2} - \left( 1 - \frac{\xi}{\mu} \right)^{1/2}
      \right] \frac{1}{2 \Delta^2} \left( 1 - \frac{\xi}{ (\xi^2 + \Delta^2)^{1/2}} \right) $
      The quantity in the square bracket can be Taylor expanded, and since
       $\left( 1 - \frac{\xi}{ (\xi^2 + \Delta^2)^{1/2}} \right) \approx
       \frac{\Delta^2}{2 \xi^2}$ for $\xi \gg \Delta$, we find
       $A_1 \approx \frac{m (2 m \mu)^{1/2}}{2 \pi^2} \frac{1}{2 \Delta}
       \left( \frac{\Delta}{2 \mu} {\rm ln} \frac{\mu}{\Delta} \right)$ in
       the $\mu \gg \Delta$ limit.

       $A_2$ is even under particle-hole symmetry.  Hence, the dominant contribution
       in the $|\Delta| \ll \mu$ limit can be approximated as
       $A_2 \approx \frac{m (2 m \mu)^{1/2}}{2 \pi^2} \int_{-\mu}^{\mu} d\xi
       \frac{1}{2E} \frac{1}{(E+\xi)^2} \to
        \frac{m (2 m \mu)^{1/2}}{2 \pi^2} \frac{1}{2 \Delta^2}$.

       For $B_1$, we substitute $\frac{k^2}{2 m} = \xi + \mu$, which generates
       respectively particle-hole asymmetric and symmetric contributions (note the extra factor
       of $\xi$ in the definition of $B_1$), and the latter one is associated with
       a large coefficient $\mu$.
       This term gives the dominant contribution to $B_1$. We obtain
       $B_1 = -\frac{2}{3} \frac{m (2 m \mu)^{1/2}}{2 \pi^2} \frac{\mu}{\Delta^2} K$
       where $K$ is the dimensionless integral
       $\Delta^2 \int_0^\mu d \xi \frac{\xi}{E^2} \frac{1}{E+\xi}
       \left(\frac{1}{E+\xi} + \frac{1}{E} \right) \to \frac{1}{2}$ for $\mu \gg \Delta$.
       Thus $B_1 \gg A$.
       We find that $B_2 \ll B_1$ in the $\mu \gg \Delta$ limit.
       (When using a constant density of states, it turns out that the particle-hole symmetric part of
       $B_2$ yields exactly zero, and hence we also need to include the energy dependence of the density of
       states.  $B_2$ is of order $(\Delta/\mu)^2$ smaller than $B_1$.)
       We obtain thus finally the dispersion
       \be
       0 = \tilde A_1 \omega + \frac{\omega^2}{\Delta} - \tilde B \frac{q^2}{m}  \label{d3}
       \ee
              where $\tilde A_1  \equiv   \frac{\Delta}{2 \mu} {\rm ln} \frac{\mu}{\Delta} $,
              $\tilde B \equiv \frac{2}{3} \frac{\mu}{\Delta}$.
       This is a quadratic equation and can be solved explicitly, but the main results
       can also be obtained by simply comparing terms in eq (\ref{d2}).
       When $ \frac{\tilde B q^2}{m \Delta} \ll \tilde A_1^2$, that is,
       when $v_F q \ll v_F q^* \equiv \Delta (\frac{\Delta}{\mu} {\rm ln} \frac{\mu}{\Delta})$,
       where $v_F \equiv \sqrt{2\mu/m}$ is the fermi velocity,
       we have

\end{widetext}

  \be
  \omega = \frac{ q^2}{2 m} \frac{ \frac{4}{3} \frac{\mu}{\Delta}}
  { \left( \frac{\Delta}{2 \mu} {\rm ln} \frac{\mu}{\Delta}
   \right)}
   \label{disp1}
   \ee

   The frequency is thus quadratic in $q$, but with a coefficient much larger
   than $1/m$.  Alternatively, we can also write
   \be
   \omega = (q \xi_0)^2 \Delta \left[ \frac{ \frac{2}{3} \pi^2}
   { \frac{\Delta}{\mu} {\rm ln} \frac{\mu}{\Delta}} \right]
   \label{disp2}
   \ee
   where
   $\xi_0 \equiv \frac{v_F}{\pi \Delta}$ is a measure  of the zero temperature coherence length.
   We note that the factor within the square bracket $[ \ ]$ is much larger than $1$.
   If $q \gg q^*$ however, the dispersion becomes linear, with
   \be
   \omega = \sqrt{ \frac{ \tilde B \Delta}{m} } q = \frac{v_F}{\sqrt{3}} q
   \label{disp3}
   \ee
   The transition between these two regime occurs at $q \approx q^*$, where
   $\omega \approx \omega^* \equiv \Delta  \frac{\Delta}{2 \mu} {\rm ln} \frac{\mu}{\Delta}$.

   The existence of this quadratic mode at small $q$ and $\omega$ can be understood
   from gauge symmetry arguments.  The equations of motion such as eq (\ref{em1})
  and (\ref{em2}) must remain valid under arbitrary gauge transformations
  (we leave out the subscripts $\vec k$ to simplify the writing here)
  $a_{\lambda} \to a_{\lambda} e^{ i \theta_\lambda}$ and
  $c_{\nu} \to c_{\nu} e^{ i \phi_\nu}$.  Hence, the equation of motion for
  $\delta \Delta_{1/2,3/2}$ must decouple from those for other $\delta \Delta_{\lambda,\nu}$ and
  also all $\delta \Delta_{\lambda,\nu}^*$'s.  For example $\delta \Delta_{1/2,3/2}^*$ transforms
  differently from $\delta \Delta_{1/2,3/2}$ under a gauge transformation of $c_{3/2}$,
  and this difference cannot be compensated by any combinations of $\Delta_{1/2,1/2}$ and
  $\Delta_{-1/2,-1/2}$.  It also cannot couple to $\delta \Delta_{-1/2,3/2}$:  for
  example $\Delta_{-1/2,-1/2}^* \delta \Delta_{-1/2,3/2}$ and
  $\Delta_{1/2,1/2}^* \delta \Delta_{1/2,3/2}$ transform in the same way under transformations
  of $a_{1/2}$ and $a_{-1/2}$, but they do not do so under transformations of
  $c_{\pm 1/2}$.  We thus obtain a scalar (not the determinant of a matrix,
   {\it c.f. (case 2) and (case 3)} below)
  equation which relates $\omega$ and $q$ (such as eq (\ref{d1})).
   The small $\omega$ and $q$ expansion of this equation
  must therefore be of the form $0=A_1 \omega + B \frac{q^2}{m}$, where $A_1$ does not
  vanish unless there is another symmetry (here particle-hole), and the expansion
  in $q$ begins with $q^2$ since our system is spatially inversion symmetric \cite{grad,nj}.
  Therefore, though our specific formulas assumed weak-coupling,  the quadratic
  dispersion at small $q$ and $\omega$ is more general, provided that
  the broken symmetries remain the same as those found within our weak-coupling
  theory.  The frequency scale $\omega^*$ unfortunately is small in
  the weak-coupling regime.  However,  we also note that
  $\omega^*$ and $q^*$ increase with $\Delta/\mu$, thus the frequency and
  momentum range where would have this quadratic mode would therer increase
  with the strength of the attractive interaction.

     As seen from eq (\ref{em1}) and (\ref{em2}), this mode couples to "spin-flip" of the c-particles.
     [This coupling is allowed because, e.g., in eq (\ref{em1}),
     $<a_{1/2}c_{3/2}>$ and $\Delta_{1/2,1/2} < c_{1/2}^{\dagger} c_{3/2} > $ transform in
     the same way under gauge transformations].  The order parameter mode under discussion
     therefore can be excited by Raman pulses which inter-convert $\nu = 1/2$ and
     $\nu = 3/2$ hyperfine sublevels.
     Bragg scattering experiments have already been performed in fermi gases \cite{bragg}.
     Though that experiment does not involve "spin-flips",
     it seems that a generalization of the method there can also observe spin-waves
     and hence the order parameter collective modes here.

     Similar discussion applies to all $\nu \ne \pm 1/2$.  There are thus $2 \times (2f -1)$ such
     modes.  The mode labeled by $\delta \Delta_{\lambda,\nu}$ couples to the observable
     $< c^{\dagger}_{-\vec k_+,\lambda} c_{-\vec k_-, \nu} >$, a generalized "spin-density".

     One can also consider $\delta \Delta^*_{\lambda,\nu}$ with $\nu \ne \pm 1/2$.  These are
     just the complex conjugates of the modes discussed above,
     (with frequencies opposite sign \cite{opp})
      and are not new physical modes.

     ({\it case 2}): $\delta \Delta_{\lambda,\nu} (\vec q)$ with $\nu = \lambda$:  We can proceed
     as above. This variable
     couples to $\delta \Delta^*_{\lambda,\nu}(-\vec q)$.  The algebra is identical with
     the ordinary BCS case for pairing between two species, and hence we would
     not display the details of our calculations here.  We just remind the
     readers that, due to the coupling between $\delta \Delta_{\lambda,\nu} (\vec q)$
     and $\delta \Delta^*_{\lambda,\nu} (\vec q)$, the dispersion is obtained
     by setting the determinant of a matrix to be zero\cite{det}
     ({\it c.f., (case 1)} above).
     In the weak-pairing limit we obtained
     the mode frequency of the Anderson-Bogoliubov mode \cite{AB} $\omega = \frac{v_F}{\sqrt{3}} q$.
     This mode can also be interpreted in a similar manner as in the two component case.
     (In the strong-pairing limit we obtain again the Bogoliubov mode for bound boson pair, similar to
     \cite{Engelbrecht}, but we shall not go into that here).  These modes couple to the densities
     fluctuations $< c^{\dagger}_{-\vec k_+,\lambda} c_{-\vec k_-, \lambda} >^{(1)}$ and
     $< a_{-\vec k_+,\lambda} a^{\dagger}_{-\vec k_-, \lambda} >^{(1)}$.
     There are two such modes, one for each choice for $\lambda$.

     ({\it case 3}) $\delta \Delta_{\lambda,-\lambda} (\vec q)$:  we find that it couples with
     $\delta \Delta^*_{-\lambda,\lambda} (-\vec q)$
     [$\Delta^*_{-\lambda,-\lambda} \delta \Delta_{\lambda,-\lambda}$ and
     $\Delta_{\lambda,\lambda} \delta \Delta^*_{-\lambda,\lambda}$ transform
     under the same way under gauge transformations.].  The equation of motion is
     analogous to ({\it case 2}).  Thus again we have linear modes with $\omega = \frac{v_F}{\sqrt{3}} q$.
     These modes couple to the (spin) densities
     $< c^{\dagger}_{-\vec k_+,\lambda} c_{-\vec k_-, -\lambda} >^{(1)}$ and
     $< a_{\vec k_+,\lambda} a^{\dagger}_{\vec k_-, -\lambda} >^{(1)}$.
     There are two such modes, again one for each choice for $\lambda$.

     The two modes in ({\it case 2}) can also be viewed as one "density" mode and one "longitudinal
     spin" mode (though the "density" mode involves only the total
     densities of the $\lambda=\nu=\pm 1/2$ particles and the "longitudinal" mode
     involves the differences between them, and the $\nu \ne \pm 1/2$ ones are not involved).
     The longitudinal spin mode and the two "transverse" spin modes in ({\it case 3})  give altogether
     three linear spin modes with $\omega = \frac{v_F}{\sqrt{3}} q$.

     We had not discussed the spin density fluctuations corresponding
     to $< c^{\dagger}_{-\vec k_+,\nu} c_{-\vec k_-,\nu'} > ^{(1)}$
     for $\nu,\nu' \ne \pm 1/2$.  They do not couple to the order parameter fluctuations.
     Within our approximation they are simply independent particle-hole pairs as in the normal state.

        \begin{widetext}

    \subsection{bosonic limit}\label{bosonic}

     For strong attractive interactions, our formalism above may not
     apply due to the appearance of multi-particle bound states.  However, to gain
     better understanding of some of our results above, it
     is instructive to consider this limit assuming we only have tightly bound
     pairs between $a$ and $c$ particles.  We would like to illuminate on the
     counting of the collective modes and the existence of quadratic versus linear
      modes.  Readers who find these points already clear are invited to
      skip this subsection.

      In this limit the system can be
     described by bosonic fields $\psi_{\lambda,\nu}$, corresponding to
     the bound state between $a_{\lambda}$ and $c_{\nu}$.    It is simple to construct
     a theory for a bosonic condensate of this system.  The Hamiltonian $H = H_K + H_{int}$ can
       be written as the sum of the kinetic part

       \be
       H_K = \int_{\vec r} \sum_{\lambda,\nu} \left[ \frac{\nabla \psi^{\dagger}_{\lambda,\nu} \psi_{\lambda,\nu}}
       {2 m_b} -\mu_b \psi^{\dagger}_{\lambda,\nu} \psi_{\lambda,\nu} \right]
       \ee
       where $m_b$ denotes the mass of the atom-pair and $\mu_b$ denotes the chemical potential,
       and the interaction, the most general form of which
     obeying SU(2)$\times$SU(6) symmetry reads

     \be
     H_{int} = \int_{\vec r} \frac{\tilde g}{8} [ \delta_{\lambda_1,\lambda_3} \delta_{\lambda_2,\lambda_4} +
       \delta_{\lambda_1,\lambda_4} \delta_{\lambda_2,\lambda_3} ]
   [ \delta_{\nu_1,\nu_3} \delta_{\nu_2,\nu_4} +
       \delta_{\nu_1,\nu_4} \delta_{\nu_2,\nu_3} ]
       \psi^{\dagger}_{\lambda_1,\nu_1}  \psi^{\dagger}_{\lambda_2,\nu_2}
       \psi_{\lambda_3,\nu_3}  \psi_{\lambda_4,\nu_4}  \ .
       \ee
       For the ground state, we replace the operators $\psi_{\lambda,\nu}$
       by c-numbers $\Psi_{\lambda,\nu}$.  The resulting
       energy reads
       \bea
       E &=& - \mu_b \sum_{\lambda,\nu} | \Psi_{\lambda,\nu}|^2
        + \frac{\tilde g}{4} \sum_{\lambda,\nu,\lambda',\nu'}
        \left\{ \Psi^*_{\lambda,\nu} \Psi^*_{\lambda',\nu'} \Psi_{\lambda',\nu'} \Psi_{\lambda,\nu}
        + \Psi^*_{\lambda,\nu} \Psi^*_{\lambda',\nu'} \Psi_{\lambda',\nu} \Psi_{\lambda,\nu'} \right\}
        \nonumber \\
        & = &   - \mu_b {\rm Tr} \left[ {\bf \Psi \Psi^{\dagger}} \right]
        + \frac{\tilde g}{4} \left\{
        \left({\rm Tr} \left[ {\bf \Psi \Psi^{\dagger}} \right]\right)^2
         + {\rm Tr} \left[ {\bf \Psi \Psi^{\dagger} \Psi \Psi^{\dagger}} \right]\right\}
        \eea       \end{widetext}
        where ${\bf \Psi}$ is a $2 \times (2 f + 1)$ matrix with entries $\Psi_{\lambda,\nu}$.
        This energy can be minimized with exactly the same procedure as in sec \ref{G}.
        We obtain $\sum_{\nu} \Psi_{1/2,\nu}\Psi^*_{-1/2,\nu} = 0$
       and $\sum_{\nu} |\Psi_{1/2,\nu}|^2 = \sum_{\nu} |\Psi_{-1/2,\nu}|^2$
       ({\it c.f.} eq (\ref{ortho}) and (\ref{eD})).
       One possible possible solution to these two equation is $\Psi_{1/2,1/2} = \Psi_{-1/2,-1/2} = \Psi$
       but with all other components zero.
       Minimizing $E$, we get $\mu_b  = \frac{3 \tilde g}{2} |\Psi|^2$.
       Again we have many degenerate ground states.

       The collective modes are also in one-to-one correspondence
       with the ones we found above in the weak-pairing regime. They can be found by
       standard Bogoliubov transformation.  We just state the results.
       The fluctuation
       $\delta \Psi_{\lambda,\nu}$ for $\nu \ne \pm 1/2$ has dispersion $\omega = \frac{q^2}{2 m_b}$,
        corresponding to free particles.
       There are $2 \times (2 f -1 )$ of these modes.
       ( $\delta \Psi^*_{\lambda,\nu}$ has $\omega = -\frac{q^2}{2 m_b}$ but simply  correspond
       to removal of a boson and are not new  modes).  These are analogous to ({\it case 1}) discussed
       above.
       The four variables $\delta \Psi_{\lambda,\lambda}$ and $\delta \Psi^*_{\lambda,\lambda}$
       with $\lambda = \pm 1/2$ are coupled (the equation determining
       the relation between $\omega$ and $q$ is then again obtained by setting the
       determinant of a matrix to zero.).  These
         yield two Goldstone modes, one with velocity
       $\omega = c q$ ($\omega = c_s q$) corresponding to in (out-of)-phase oscillations
       between the $\lambda = 1/2$ and the $\lambda = -1/2$ components.
       ($\delta \Psi_{1/2,1/2} = + (-) \delta \Psi_{-1/2,-1/2}$ and
       $\delta \Psi^*_{1/2,1/2} = + (-) \delta \Psi^*_{-1/2,-1/2}$).
       We obtain $c_s = \frac{1}{\sqrt{3}} c$ and $c^2 = \mu_b/m_b$.
       These are the analogous to the modes in ({\it  case 2}).  Finally,
       the four variables $\delta \Psi_{\lambda,-\lambda}$ and $\delta \Psi^*_{\lambda,-\lambda}$
       with $\lambda = \pm 1/2$ yield two other Goldstone modes again with $\omega = c_s q$.
       These are the analog of ({\it case 3}).
       We have thus in total one "density" mode and three "spin" modes as discussed above.
       In the present case however the "density" modes and "spin" modes
       have different velocities due to interactions among bosons.
       The finding that the "density" and "spin" modes have the same velocities
       (as well as that they are given by $v_F/\sqrt{3}$)
       in the weak-coupling limit in the last subsection is the result of the approximation employed there,
       and is not expected to hold in general.

       \section{Comparison with SU(N)}\label{Comp}

       We now compare our results with BCS pairing in SU(N) models , where
       there are $N$ species of fermions $a_{1,...,N}$ with interspecies interaction
       which is invariant under $SU(N)$ transformations.
       We have seen that, for our $SU(2) \times SU(6)$ system with interspecies
       pairing, for the {\em ground state}, the pairing order parameter
       reduces to, with a suitable choice of basis, one where each $a$-species
       pairs only with one unique $c$-species.  We have seen that this
       is a consequence of eq (\ref{ortho}).  The results for SU(N) are similar\cite{Modawi97,HH}
       but there is one important difference.  For SU(N), the order parameter ${\bf \Delta}$ is
       an antisymmetric $N \times N$ matrix (there is no such restriction
         for our SU(2)$\times$ SU(6) case).   Under a unitary transformation ${\bf U}$,
       it transforms as ${\bf \Delta} \to {\bf U \Delta U^t}$ where the superscript
       $t$ denotes the transpose.  Hence, with a suitable ${\bf U}$,
       ${\bf \Delta}$ can always be transformed \cite{Youla}
       to one where all entries $\Delta_{\lambda_1,\lambda_2}$ vanish except
       \be
       \Delta_{12}=-\Delta_{21} \qquad \Delta_{34}=-\Delta_{43} \qquad ... \ ,
       \label{anti1}
       \ee
       that is,
       \be
       \Delta_{13}=\Delta_{14}= ... = \Delta_{23}= ... = 0
       \label{anti2}
       \ee
       so that $1$ only pairs with $2$, $3$ only pairs with $4$ etc \cite{extension}.
       However, we emphasize that the origin
       of eqs (\ref{anti1}) and (\ref{anti2}) is very different from the SU(2)$\times$ SU(6) case.
       The possibility of writing ${\bf \Delta}$ in  the form of Eqs (\ref{anti1}) and (\ref{anti2})
       is purely a consequence of the fact that ${\bf \Delta}$ is antisymmetric \cite{Youla},
       which in turn only requires
       fermionic anticommutation relations, and therefore
       holds for {\em arbitrary} states (including excited states) for the system.
       For our $SU(2)\times SU(6)$ system,  eq (\ref{ortho}) needs not hold other than
       the ground state.

      Now we compare the collective modes.   Consider first $N=3$.
       In the gauge where the only finite
      component is $\Delta_{12}$, it can easily be shown that $\delta \Delta_{13}$
      obeys the same equations of motion as eq (\ref{em1}) and (\ref{em2})
      with $a_{1/2} \to a_1$, $c_{1/2} \to a_2$, $c_{3/2} \to a_3$.  Hence
      the  dispersion found there also applies.  That $\delta \Delta_{13}$ decouples
      from other order parameter modes can also be seen by gauge invariance
      arguments as in sec \ref{Modes}.  It can be easily seen that $\delta \Delta_{23}$
      yields yet one more mode with the same dispersion as $\delta \Delta_{13}$.
      On the other hand, $\delta \Delta_{12}$ and $\delta \Delta^*_{12}$ are
      coupled together, and they generate one linear mode as in {\it case 2}
      in sec \ref{Modes}.  Hence in total,  we have one linear mode, and
      two modes with quadratic dispersions for small $\omega$ and $q$.

      Our results for the number of linear and quadratic modes agree with
      Ref \cite{He06}.  (note that they have an alternative but related argument
      for the existence of quadratic modes).  However, the precise
      dispersion of the quadratic mode is different. They obtain
      a dispersion same as a free particle of mass $m$.  Our eq (\ref{d1})
      is the same as their (A1) at zero temperature (except $\vec k \to \vec k - \frac{\vec q}{2}$,
      see however \cite{convg}), so we believe  that they have made a subsequent
      algebraic error (the contributions from our $B_1$ and $B_2$ terms
      seem to be missing).   That their result is unreasonable was also
      pointed out in Ref. \cite{CY08} (see also \cite{grad}).   If particle-hole asymmetric terms
      are ignored, we saw that the quadratic modes become linear with velocity
        $v_F/\sqrt{3}$, in agreement with Ref. \cite{CY08}, resulting in therefore
        three linear modes in total.
        Ref. \cite{CY08} however did not provide the full dispersion relation in the presence
        of particle-hole asymmetric contributions.  Also, we do not understand their claim
        that, when particle-hole asymmetric terms are included as we have
        done here, "two of the massless modes split into a massless mode with
        quadratic dispersion and a massive one".  We have seen that $\delta \Delta_{13}$ and
        $\delta \Delta_{23}$ yield two independent quadratic modes.  These two
        modes would not couple to each other, nor to other order parameter
        modes by gauge symmetry, and thus each must yield one quadratic mode.
        Since this is due to gauge symmetry,
        this result should hold even beyond the weak-coupling approximation we have employed.
        Lastly, Ref. \cite{HH} has counted the number of modes differently from here
        (and therefore also Ref. \cite{He06}).  They regard $\delta \Delta_{13}^*$ and
        $\delta \Delta_{23}^*$ as giving two other modes in addition to
        $\delta \Delta_{13}$ and $\delta \Delta_{23}$, and they therefore counted
        in total five Goldstone modes.  As remarked before, $\delta \Delta_{13}^*$ and
        $\delta \Delta_{13}$ just correspond to annihilation and creation of  the
        same excitation,  so it seems natural not to count them
        separately (see subsection \ref{bosonic}).  Also, Ref. \cite{HH} has
        performed a numerical calculation of the dispersion for the mode
        $\delta \Delta_{13}$.  Their numerical result seems only to show
        a linear dispersion, likely due to the small $\Delta/\mu$ values chosen there.

        The above discussions can readily generalized to larger N.
        For example, for SU(5), there are four quadratic modes.
        With choice of order parameter as in eq (\ref{anti1}) and (\ref{anti2}), they
        are $\delta \Delta_{\lambda,5}$ for $\lambda = 1,2,3,4$.
        No such modes are expected for $N$ being even.
        For SU(4), there are six linear but no quadratic modes.

      \section{Conclusion} \label{Concl}

   In this paper, we have considered some superfluid properties of
   an SU(2)$\times$SU(6) system with interspecies pairing, motivated by
   the system studied experimentally in ref \cite{Taie}.   We considered
   both the ground state and collective excitations.  Some properties
   are dramatically different from two-component systems. There are in particular
   collective mode excitations with quadratic dispersions at low frequencies.
   Many of our results found are generally applicable to systems with interspecies pairing
   with high symmetries.

   \section{Acknowledgement}

 The author would like to thank Miguel Cazalilla for drawing his attention to
 ref \cite{Taie}, and Feng-Kuo Hsu for noting an algebraic error in the first
 version of this manuscript.
 This research was supported by the  National Science Council of
 Taiwan.


\begin{thebibliography}{plain}
\bibitem{D2}
 Eq (\ref{ortho}) and (\ref{eD}) are  applicable also to
 the case where the pairing occurs in a fixed total spin $F$ channel
  where $F = f+\frac{1}{2}$
 or $F= f - \frac{1}{2}$.  In this case, we can write
 $\Delta_{\lambda,\nu} = \sum_M
 \Delta_{FM} < \frac{1}{2} f, FM|\frac{1}{2} f, \lambda, \nu>$
 where $ < \frac{1}{2} f, FM|\frac{1}{2} f, \lambda, \nu>$ is the
 Clebsch-Gordan coefficent of combining the states
 $|\frac{1}{2},\lambda>$ and $|f,\nu>$ to $|FM>$.  Note
 that $F$ is not summed.  Substituting this into eq (\ref{ortho}) and (\ref{eD})
 and using the values for the Clebsch-Gordan coefficients,
 we recover the results of \cite{Dickerscheid}
  that the state is "non-magnetic".
\bibitem{convg} Note that we used $\vec k \pm \vec q/2$ for the operators $a$ and $c$.
This guarantees that there would not be terms which
behave as ${\vec k \cdot \vec q }/{k^4}$ for $k \to \infty$
which would converge only conditionally.
\bibitem{grad}  The term $(\frac{A_1}{2} + B_1 + B_2) \frac{q^2}{m}$ in eq (\ref{d2}) can be understood
as the (negative) of a gradient energy in the presence of $\delta \Delta_{1/2,3/2}(\vec q)$,
where we have the extra terms in the Hamiltonian
given by $\delta H = \sum_{\vec k} \left[ \delta \Delta_{1/2,3/2}(\vec q)
a^{\dagger}_{\vec k_+,1/2} c^{\dagger}_{-\vec k_-, 3/2} +
\delta \Delta^*_{1/2,3/2}(\vec q)
c_{\vec k_-,3/2} a_{-\vec k_+, 1/2} \right] $
$- \frac{ |\delta \Delta_{1/2,3/2}(\vec q)|^2}{g}$.  The change of
energy $\delta E$ of the system to second order in $\delta \Delta_{1/2,3/2}(\vec q)$,
can be evaluated by perturbation theory.  We get
$\delta E = \kappa  |\delta \Delta_{1/2,3/2}(\vec q)|^2$ where
$\kappa$ is simply negative of the expression in the curly brackets in eq (\ref{d1})
with $\omega = 0$.
\bibitem{nj}  An alternative way to understand eq (\ref{d2}) is to note
that one must have a continuity equation for the spin density
$ \delta n_s \equiv \sum_{\vec k} <c^{\dagger}_{-\vec k_+,1/2} c_{-\vec k_-, 3/2}>^{(1)}$
in the form $ \omega \delta n_s - \vec q \cdot \vec J_s = 0$,
where $\vec J_s$ is a spin current.  Indeed, this equation can be obtained
by summing eq (\ref{em2}) over all $\vec k$.
$- \vec q \cdot \vec J_s$ can be written as, making use of the gap equation eq (\ref{sc1}),
$- \vec q \cdot \vec J_s =$
$\sum_{\vec k} (\xi_{k_+} - \xi_{k_-}) < c^{\dagger}_{-\vec k_+,1/2} c_{-\vec k_-, 3/2}>^{(1)}$
$+ \Delta^*_{1/2,1/2} \delta \Delta_{1/2,3/2} ({\vec q})
\sum_{\vec k} \left[ \frac{1}{2 E_{k_+}} - \frac{1}{2 E_k} \right]$
where the last term arises from the modification of the current operator
in the presence of $\delta H$.  The distribution function
$<c^{\dagger}_{-\vec k_+,1/2} c_{-\vec k_-, 3/2}>^{(1)}$ can be
evaluated from eq (\ref{em1}) and (\ref{em2}) or directly from perturbation theory.
We have
$ <c^{\dagger}_{-\vec k_+,1/2} c_{-\vec k_-, 3/2}>^{(1)} = $
$ \frac{1}{2 E_{k_+}} \left[
\frac{ 1 - f(\xi_{k_-})}{ E_{k_+} + \xi_{k_-} - \omega} -
\frac{  f(\xi_{k_-})}{ E_{k_+} - \xi_{k_-} + \omega}  \right]
\Delta^*_{1/2,1/2} \delta \Delta_{1/2,3/2}(\vec q)$.
Expanding in small $q$ and $\omega$, we obtain
$ \delta n_s = [ A_1 + A_2 \omega + ... ] \Delta^*_{1/2,1/2} \delta \Delta_{1/2,3/2}(\vec q)$
and
$ - \vec q \cdot \vec J_s =
[(\frac{A_1}{2} + B_1 + B_2) \frac{q^2}{m} + ...] \Delta^*_{1/2,1/2} \delta \Delta_{1/2,3/2}(\vec q)$
where $ ...$ denote terms of higher order.  The continuity equation
then implies eq (\ref{d2}).  Hence $A_1$ represents the static susceptibility
of the spin density $\delta n_s$ in response to the perturbation $\delta \Delta_{1/2,3/2}(\vec q)$.
Note that
$ - \vec q \cdot \vec J_s $ is therefore related to the gradient energy discussed in \cite{grad}.
\bibitem{opp} The dispersion is given by eq (\ref{d2}) with $\omega \to - \omega$.
\bibitem{det} This equation has the form \cite{Engelbrecht}
\bdm
\left|
\ba{cc} M + B'_1 q^2 + A'_1 \omega + A'_2 \omega^2 + ...  &
M + B'_2 q^2 + C'_1 \omega + ... \\
M + B'_2 q^2 - C'_1 \omega + ...&
M + B'_1 q^2 - A'_1 \omega + A'_2 \omega^2 + ...
\ea
\right|
 = 0
\edm
  Here $M, B'_1, B'_2, A'_1, A'_2, C'_1$ are
coefficients and $...$ denotes higher order terms in  $q$ and/or $\omega$.
\end{thebibliography}
\end{document}